# A Mathematical Theory of Redox Biology

James N. Cobley[1] and Michalis G. Nikolaidis[2]

1. The University of Dundee, Dundee, Scotland, UK.
2. Aristotle University of Thessaloniki, Serres, Greece.

**Correspondence**: (jcobley001@dundee.ac.uk) or (j_cobley@yahoo.com)

**ORCID**: https://orcid.org/0000-0001-5505-7424 (James) and https://orcid.org/0000-0001-6165-8437 (Michalis)

**Abstract**
Redox biology underpins signalling, metabolism, immunity, and adaptation, yet lacks a unifying theoretical framework capable of formalising structure, function, and dynamics. Current interpretations rely on descriptive catalogues of molecules and reactions, obscuring how redox behaviour emerges from constrained biochemical organisation. Here, we present a mathematical theory of redox biology that resolves this gap by treating redox systems as finite, compositional, dynamical, and spatially embedded objects. We define a structured redox state space in which admissible molecular transformations form a neutral algebra of possibilities. Biological function emerges when this structure is embedded within a wider molecular network and interpreted through weighted flux distributions. Time-dependent reweighting of these transformations generates redox dynamics, while spatial embedding enforces locality and causality, yielding a distributed redox field. Within this framework, context dependence, nonlinearity, hysteresis, and memory arise naturally from bounded state spaces and irreversible transformations, without requiring ad hoc assumptions. This theory provides a working, predictive interpretative basis for redox biology: it constrains admissible states and trajectories, clarifies the meaning of redox measurements, and links chemical transformation to biological behaviour. Redox biology emerges as a geometric, dynamical process governed by lawful organisation.

**1. Introduction**
Redox biology is now widely recognised as essential for life, playing a central role in signalling, metabolism, immunity, adaptation, and the biological basis of health and disease across the lifespan [1–10]. Experimentally, the field has advanced to the point where detailed measurements of small molecules, protein residues, and pathway-level responses can be made across conditions, time, and space [11–18]. These measurements, however, probe not isolated molecules but local chemical environments—structured contexts in which redox reactions occur, interact, and propagate. Despite this experimental maturity, redox biology lacks a theoretical framework capable of formalising the structure, function, and dynamics of these environments [19]. As a result, data interpretation relies on descriptive inventories and narrative integration rather than on the principled unification of experiment with theory [20,21].

Objects of study in redox biology are typically organised through descriptive molecular catalogues, such as the grouping of small molecules into lists including "reactive species", "reactive oxygen species", or "reactive sulfur species". While useful heuristics [22–24], these catalogues abstract molecules from the environments in which they act and do not expose the underlying structure of redox systems [25]. Similarly, grouping molecules by functional action—such as "oxidant", "antioxidant", or "reductant"—assigns context-dependent roles as if they were invariant properties. Even though the same molecular entity can act in multiple, and opposing, ways depending on its local environment and reaction partners [26]. These representations succeed at description but fail to define a coherent state space for the environment itself, obscuring how objects relate, compose, and transform.

Redox function is an emergent, context-dependent property of this environment. Functional properties are determined relationally by how local molecular transformations interact within a dynamic network, rather than being intrinsic and invariant labels (e.g., "good"). In this framing, redox biology is naturally understood as a spatially embedded, dynamic interaction network, in which biological function emerges from the configuration and evolution of redox transformations across space and time [27,28]. Function is therefore relative and not fixed, free to change as the structure and dynamics of the environment evolve. Despite this implicit network-dynamical and geometric nature, redox biology currently lacks a mathematical formalism capable of representing the environment as such an object. Without such a framework, structure, state, and dynamics remain implicit, and functional interpretation relies on narrative reasoning rather than derivation from a defined state space.

In this work, we formalise redox biology as a geometrically embedded dynamical network of local transformations. We first define a minimal compositional structure for small-molecule redox chemistry (the Core Redox Module, CRM) as a symmetric monoidal category of objects and morphisms. Here, "symmetric monoidal" simply means that redox reactions can be composed in parallel and that the order of independent reactions does not matter. We then embed the CRM into a wider cellular interaction graph that includes a Modifying Redox Module (MRM) and the redoxome, and define redox state as a functorial readout of weighted flux on this graph. Next, we extend this mapping to dynamics by evolving edge weights in time under a constrained flow law. Finally, we embed the resulting algebra into physical space as a field of local categories coupled by transport, yielding a redox field theory that unifies structure, function, and dynamics under locality and causality (see Box 1 for a plain-language theoretical guide).

> **Box 1. A plain-language guide to the mathematics behind the theory of redox biology**
>
> The mathematics in this paper is not about solving equations for single reactions. Instead, it provides a way to describe **what redox chemistry can do, how it combines, and how biological meaning emerges from its use**.
>
> The first step is structural. The theory treats key redox-active molecules as belonging to a **finite set of physically allowed molecular states**, and it treats every chemically possible transformation between those states—such as oxidation, reduction, proton transfer, or excitation—as an allowed process linking one state to another. This turns redox chemistry into a **structured network of states and transformations**, rather than an unorganised list of reactions.
>
> Crucially, this network is **compositional**. Individual transformations can be chained into pathways, or they can occur independently and at the same time. The mathematics used to formalise this comes from **category theory**, which is a precise language for describing how processes connect, combine, and remain independent when they should. This does *not* assume linearity or equilibrium; it simply encodes which transformations are physically admissible and how they may compose.
>
> Biological redox state is then defined not as a label like "oxidised" or "reduced", but as **how this network is being used**. Over a given time window, each possible transformation is assigned a non-negative weight that represents how active that process is under the conditions studied. Together, these weights describe a pattern of activity across the network—what reactions are carrying flux and where.
>
> To make this pattern comparable across conditions, tissues, or experiments, the theory introduces a **projection step**. The distributed activity on the network is condensed into a compact numerical summary that reflects the balance of redox-relevant channels. This step uses familiar tools from linear algebra, but only as a *readout*: it summarises network activity without changing the underlying structure or dynamics.
>
> The theory then becomes dynamic by allowing the pattern of activity to **change over time under constraints**. Regulation, enzyme availability, and cellular context do not create new reactions; instead, they reshape how flux is routed through existing ones. Because these constraints act on how the system moves rather than where it is, the resulting dynamics can be history-dependent: what the system does next can depend on how it arrived there. Finally, the model is embedded in physical space. Instead of treating a cell or tissue as a single well-mixed container, the theory treats it as many neighbouring microscopic regions, each running its own local version of the same redox network. These regions are coupled because molecules and signals can move between them. As a result, redox state becomes a **spatially distributed field** that can form gradients, propagate locally, and evolve over time—much like temperature or electrical potential—while still respecting causality and physical locality.

> In short, the theory provides a way to move from *what redox chemistry allows*, to *how it is used*, to *how biological meaning emerges*, without reducing redox biology to either static reaction lists or simple equilibrium variables.

## 2. Results and Discussion

This work establishes a formal mathematical theory of redox biology. Each section introduces a precise mathematical object or dynamical law. The primary contribution is theoretical: we define the structure, function, dynamics, and spatial embedding of redox systems in a unified framework. To aid biological intuition, examples and interpretative discussion sections are included alongside each part of the formal framework. In addition, plain language summaries of both the theory and technical terms are provided (**Box 1 and 2**). Since formal derivations are provided (**Supplemental Notes**), the main text presents the core construction required to interpret redox biology as a structured, dynamical system.

> **Box 2. A plain-language glossary of key mathematical and physical terms**
>
> **Admissible**
>
> Allowed by physics and chemistry. An admissible transformation is one that can actually occur under the rules of the model, not merely one that can be imagined or written symbolically.
>
> *Under these conditions, this one reaction is admissible, whereas another is not.*
>
> **Composition**
>
> The rule for combining transformations. Transformations may occur in sequence (forming pathways) or independently in parallel. Complex behaviour is built by composing simpler processes without introducing new chemistry.
>
> *Two admissible steps can be composed to take the system from A to C via B, even if no single direct step A → C exists.*
>
> **Constraint**
>
> A restriction on how states or activities may be occupied or evolve. Constraints limit what can happen in practice without changing which transformations exist.
>
> *Non-negativity constrains how flux weights can evolve, even though the underlying transformations remain admissible.*
>
> **Domain**
>
> The physical region being modelled, such as a cell, organelle, or tissue patch. The domain determines which molecules are present and which transformations can be expressed locally.
>
> *Changing the domain from whole cell to mitochondrial matrix changes which redox processes are relevant.*
>
> **Dynamics / Dynamical**
>
> How the system changes over time. A dynamical description treats redox biology as an evolving process rather than a static catalogue of reactions.
>
> *The same redox network can support different behaviours as activity is redistributed over time.*
>
> **Edge / Morphism**
>
> An arrow representing an allowed transformation between states. In network language this arrow is called an edge; in category-theoretic language the same arrow is called a morphism.

> *The conversion of hydrogen peroxide to water is an edge in the network and the same arrow is a morphism in the category.*
>
> **Finite**
> Explicitly limited in size. Only a bounded set of states and transformations is included, ensuring the system remains physically meaningful and computationally tractable.
> *The CRM uses a finite state space rather than an open-ended reaction universe.*
>
> **Flux**
> A quantitative measure of how much activity passes through a transformation over a chosen time window. Flux reflects usage, not mere possibility.
> *Flux weights identify which transformations are actively utilised under given conditions.*
>
> **Mapping**
> A rule that converts one description into another in a consistent way. Mappings are used to relate network activity to redox state summaries or biological readouts.
> *A mapping can project distributed network activity into a compact redox state profile for comparison across conditions.*
>
> **Matrix**
> A structured table of numbers that acts as a conversion rule, translating one set of quantities into another in a reproducible way. In this framework, matrices summarise how network structure relates activity on transformations to state-level descriptions.
> *The incidence matrix encodes which transformations connect to which states, allowing network activity to be condensed into a redox state vector.*
>
> **Module**
> A coherent sub-part of the system that can be treated as a unit. Modules can be combined to build larger systems without redefining their internal structure.
> *Separating the CRM from the MRM allows kinetics to change without changing which transformations exist.*
>
> **Network**
> The full web of states and transformations. A network representation captures system-level organisation rather than isolated reactions.
> *The same network can support buffering, signalling, or damage depending on how activity is routed.*
>
> **Object**
> The entities that transformations act on. In this framework, objects represent redox-relevant molecular states or state classes. Objects are connected by edges or morphisms.
> *Hydrogen peroxide is an object with multiple competing outgoing transformations.*
>
> **Redox field**
> A redox state defined continuously across space and time. Each location carries its own local redox activity, coupled to neighbouring locations by transport.
> *A redox field can exhibit spatial gradients, such as a more reduced cytosol alongside a more oxidised mitochondrial region.*
>
> **Spatial embedding**
> Placing the abstract redox network into physical space, so that location, transport, and locality matter.
> *A transformation may exist in principle, but if the relevant molecules or enzymes are elsewhere, it will not occur locally.*
>
> **State space**
> The complete set of allowed molecular states and transformations being modelled,

> together with explicit physical bounds on their occupancy.
> *A perturbation changes where the system is within state space, not what the state space itself is.*
>
> **Time-dependent**
> Explicitly changing with time. Fluxes, inputs, or activities are time-dependent if their values vary during a process.
> *During a stimulus, some transformations increase in activity while others decrease, even if concentrations appear similar.*
>
> **Transformation**
> An allowed process that converts one state into another within the network.
> *Oxidation, reduction, excitation, and protonation are transformations.*
>
> **Weight**
> A non-negative number attached to a transformation that summarises how active it is under specific conditions. Weights allow the same network structure to represent different physiological contexts.
> *Changing weights reroutes activity without changing the underlying chemistry.*

## 2.1. Structure—Defining the Redox State Space
### 2.1.1 Claim: Redox chemistry admits a finite, compositional state space

Redox biology is fundamentally concerned with a restricted set of small, diffusible molecules—such as oxygen, hydrogen peroxide, nitric oxide, and related species—and the biochemical transformations that interconvert them [29–31]. These molecules and reactions are typically represented as unstructured catalogues or reaction lists. Such representations succeed at enumeration but fail to expose the underlying structure of the system: how redox entities relate, compose, and transform within a bounded biochemical space.

We claim that redox biochemistry admits a finite, well-defined mesoscopic state space, whose elements are discrete molecular configurations representing coarse-grained equivalence classes of underlying Boltzmann microstates, and whose transitions are governed by physically realisable transformations per the laws of quantum chemistry [32]. Here, "state space" denotes the set of admissible molecular types and configurations; quantitative abundances and fluxes over this space are introduced later when dynamics are defined [33].

The state space is compositional: elementary redox transformations can be chained, combined, or occur in parallel, giving rise to higher-order pathways and networks [34]. Any formal theory of redox biology must therefore capture not only which states exist, but how transformations compose.

### 2.1.2 Construction: The Core Redox Module as a categorical object

To formalise this structure, we define the <u>C</u>ore <u>R</u>edox <u>M</u>odule (CRM) as a mathematical object that encodes both redox-relevant molecular states and the transformations between them. In the CRM, each small-molecule redox entity is treated as an object, and each biochemically admissible transformation—such as oxidation, reduction, excitation, protonation, or deprotonation [35]—is treated as a morphism linking one object to another. Each morphism represents a physically realisable transformation, admissible under the laws of quantum mechanics, statistical mechanics, and thermodynamics.

This organisation naturally takes the form of a directed graph enriched with algebraic structure. Nodes correspond to discrete molecular states (e.g., ground state dioxygen [36]), while edges represent transformations that map one state to another. These edges define operations that can be composed sequentially or occur concurrently.

To capture this compositionality explicitly, the CRM is formalised as a symmetric monoidal category. Sequential composition represents reaction chains—such as the stepwise four-electron reduction of ground-state dioxygen to water—while the monoidal (tensor) product represents parallel or independent transformations occurring within the same biological context. This construction allows redox chemistry to be treated as an algebra of transformations.

Not all CRM morphisms correspond strictly to electron-transfer reactions. Transformations involving energy exchange or coupling to external fields—for example, the photoexcitation of ground-state oxygen to singlet oxygen [37]—are included by treating external agents (such as photons) as additional objects participating in composite morphisms. In this way, the CRM unifies electron, proton, and energy transfer within a single formal structure.

The full categorical definition, including object states, morphism typing, and compositional rules, is provided (**Supplemental Note 1**). While the CRM defines the space of admissible molecular states and transformations, the realised biological state additionally depends on bounded molecular abundances within physical volumes (**Section 2.3**). The key point is that: the CRM defines a finite, closed, and compositional redox state space.

### 2.1.3 Molecular composition of the CRM
The categorical definition of the CRM is independent of any specific choice of molecular objects. To apply this structure, we instantiate the CRM with a minimal working set of small, diffusible redox-active molecules (e.g., superoxide anion) that are broadly recognised to be central to redox biology [38].

This set is heuristic rather than exhaustive. Its purpose is to define a tractable state space on which the formal structure operates, not to enumerate all molecular entities capable of redox chemistry. Inclusion is therefore based on physical diffusion range, participation in electron/proton/energy transfer, and relevance across multiple biological contexts [22].

Under this instantiation, canonical families such as reactive oxygen and nitrogen species form overlapping subsets of the CRM object set [23]. Entities requiring a macromolecular scaffold for stability or identity (e.g., a lipid peroxyl radical [38]) are excluded at this level and incorporated at higher layers of organisation.

### 2.1.4 Example
To instantiate the CRM formalism, we consider a canonical subset of reactive oxygen species [39]. Each element of this subset defines a discrete CRM object corresponding to a mesoscopic molecular identity. For example, the superoxide anion constitutes a single CRM object that represents a many-to-one coarse-graining of underlying quantum mechanical and Boltzmann microstates, conditioned on its local chemical environment (**Fig. 1A**). This object does not encode explicit electron configurations but instead defines an equivalence class of microstates that are indistinguishable at the level of redox chemistry.

Each CRM object admits a set of invariant descriptors that characterise its physical identity. For the superoxide anion, these include the presence of a single unpaired electron (free-radical character), a net negative charge, and a fixed elemental composition comprising two oxygen atoms [40]. These descriptors define the object's position within the CRM state space and determine which transformations are physically admissible. A physically admissible transformation is represented by a morphism. For example, the univalent reduction of ground-state molecular dioxygen to the superoxide anion (**Fig. 1B**).

Structurally, the reactive oxygen species subset of the CRM is organised as a directed graph, in which each CRM object corresponds to a node and each edge represents a physically admissible transformation mapping one object to another in the redox state space (**Fig. 1C**). For example, ground-state molecular dioxygen and the superoxide anion are connected by a directed edge (morphism) representing univalent reduction. Algebraically, these admissible morphisms form a symmetric monoidal category, allowing transformations to compose both sequentially and in parallel (**Fig. 1D**).

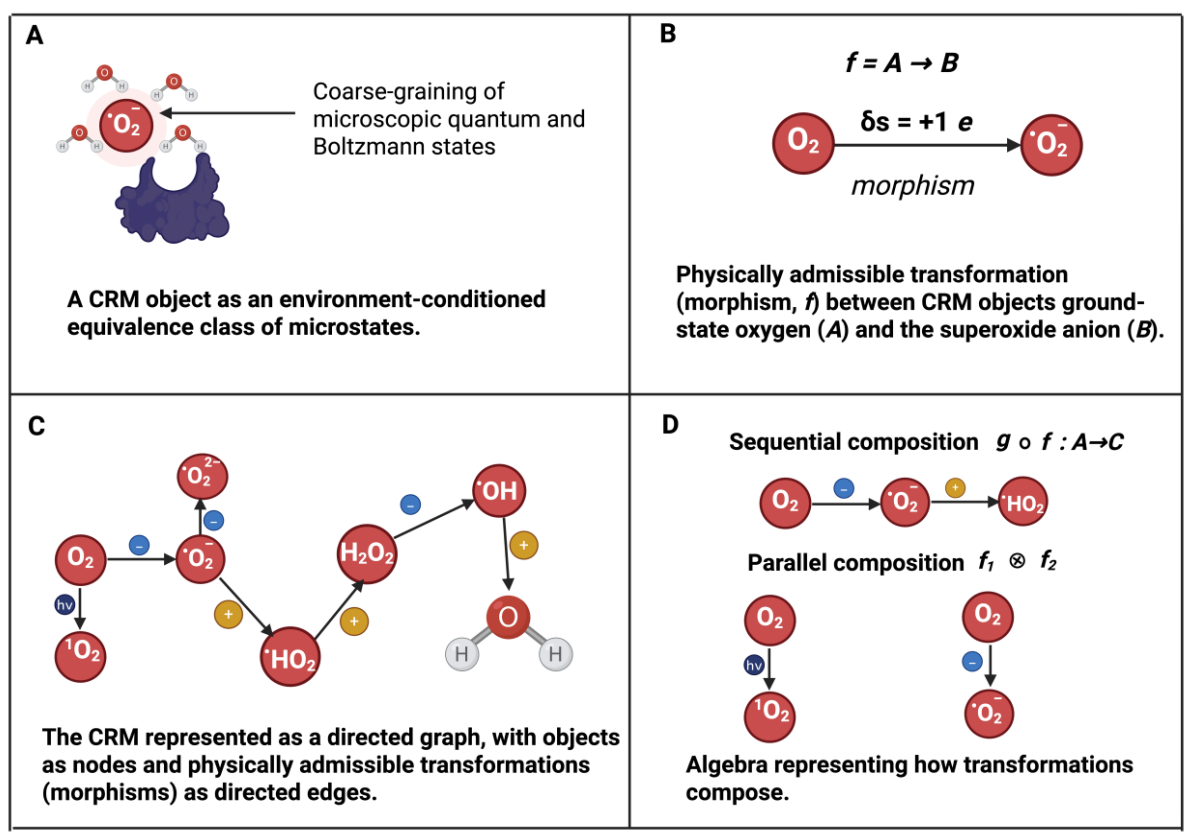

**Figure 1 | Structural foundations of the Core Redox Module (CRM)**
(A) A CRM object represents an *environment-conditioned mesoscopic equivalence class* of microscopic quantum and Boltzmann microstates. Individual molecular identities (e.g., the superoxide anion) are treated as coarse-grained state classes that implicitly encode coupling to the surrounding chemical environment rather than explicit electron-level configurations.
(B) A physically admissible transformation between CRM objects is represented as a *morphism* $f: A \to B$. Here, reduction of ground-state oxygen to the superoxide anion is shown as a state transition characterised by a change in the state descriptor $\delta s = +1\,e$. Morphisms encode lawful transformations permitted by quantum mechanics, statistical mechanics, and thermodynamics.

(C) The CRM instantiated as a directed graph. Objects correspond to discrete redox-active molecular state classes (nodes), and directed edges correspond to physically admissible transformations (morphisms), including electron transfer, proton transfer, and excitation. Canonical reactive oxygen species and their interconversions form a structured, compositional redox state space.

(D) Algebraic composition of CRM morphisms. Sequential composition $g \circ f$ represents reaction chains or pathways, while parallel (monoidal) composition $f_1 \otimes f_2$ represents concurrent, independent transformations occurring within the same biochemical context. Together, these operations define the compositional algebra underlying redox chemistry.

### 2.1.5 Interpretation: Why neutrality matters biologically

The CRM is functionally neutral. Molecular entities are defined by physical state descriptors—such as protonation state—not by biological labels or inferred roles. Likewise, morphisms denote permissible transformations without prescribing their biological meaning or consequence [29].

Neutrality is essential. In biological discourse, molecules are often assigned invariant functional labels—"oxidant", "antioxidant", "toxic", or "protective". However, the same molecular entity can act in opposing roles depending on context [41]. For example, hydrogen peroxide may be metabolised harmlessly, participate in regulated signalling, or contribute to damaging biochemistry, depending on how it is embedded in the surrounding network of reactions [42,43].

By separating what transformations are possible from how they are used, the CRM avoids conflating structure with function. Biological meaning does not reside in individual molecules or reactions but emerges only when these neutral elements are embedded in a larger relational context. Structure therefore precedes function: the CRM defines the space of possibilities, while function arises from how that space is traversed.

In this way, the CRM replaces descriptive catalogues with a principled state space. It provides the structural foundation upon which function, dynamics, and spatial organisation can later be defined, without imposing semantic bias at the level of molecular identity. This neutral, compositional framing is the first step in this mathematical theory of redox biology.

## 2.2 Function — Emergence from Network Embedding
### 2.2.1 Claim: Redox function is an emergent, relational property

Redox function is not an intrinsic property of individual molecules or reactions. Instead, biological function emerges from how redox entities are embedded within a broader molecular network and how transformations are distributed across that network. The same molecular species can participate in signalling, homeostatic regulation, or damaging biochemistry depending on context, connectivity, and constraint [44].

Accordingly, redox function cannot be assigned at the level of isolated objects within the CRM. While the CRM defines what transformations are possible, it does not determine how those transformations are realised or prioritised in living systems. Function arises only once the CRM is embedded into the wider biochemical environment, where enzymes, cofactors, and macromolecular targets modulate, channel, and interpret flux.

We therefore claim that redox function is an emergent, relational property: it is defined by patterns of connectivity and transformation across a structured molecular network.

### 2.2.2 Construction: Embedding the CRM into a functional redox network

To formalise the emergence of function, the CRM is embedded within a larger composite network comprising three interacting components:

1. the CRM, defining the space of admissible small-molecule redox transformations;
2. the <u>M</u>odifying <u>R</u>edox <u>M</u>odule (MRM), consisting of enzymes, cofactors, and protein systems that mediate, catalyse, or constrain these transformations; and
3. the redoxome, encompassing all remaining molecular entities capable of participating in redox interactions.

Together, these components form a directed, typed network in which nodes represent molecular states and edges represent transformations. Each edge is associated with a flux weight, representing the effective exchange of electrons, protons, or energy along that transformation over a defined observation window. The resulting network therefore encodes both structure (which transformations exist) and state (which transformations are active and to what extent).

Within this framework, redox function is defined as a mapping from network structure and flux distribution to biological outcome. Formally, this is expressed as a structure-to-function mapping that aggregates local redox transformations into global or subsystem-level redox states, which can then be related to biological readouts such as signalling activity, metabolic control, or stress responses [45].

The mathematical definition of this mapping—treating redox state as a function of weighted network connectivity is provided (**Supplemental Note 2**). The key point is that: function is not assigned to nodes or edges individually, but to patterns of flux across the entire network or across subgraphs.

### 2.2.3 Composition of the MRM and redoxome

*A Mathematical Theory of Redox Biology*

The MRM comprises finite classes of objects (i.e., proteins and cofactors) and morphisms (transformations) that together modify or realise CRM transformations (**Table 1**). These objects are treated as discrete mesoscopic molecular entities with their own state spaces, such as the proteoform state space [19]. Each MRM object is defined by its mode of interaction with the CRM.

**Table 1. Subgroup composition of the MRM.**

| MRM subgroup | Objects (abbreviation) | Comment/justification |
|---|---|---|
| **Frontline enzymes** | Superoxide dismutase (SOD) isoforms; Catalase; Peroxiredoxin (PRX) isoforms; Glutathione peroxidase (GPX) isoforms | Direct catalytic interface with CRM morphisms. Control formation and removal of ROS objects (e.g., $O_2^{\bullet-}$, $H_2O_2$). |
| **Frontline small molecules** | Ubiquinone / ubiquinol species; Vitamin E (α-tocopherol / tocopheroxyl radical); Vitamin C (ascorbate / dehydroascorbate); Iron and copper redox couples ($Fe^{2+}/Fe^{3+}$, $Cu^+/Cu^{2+}$); Glutathione (GSH/GSSG) | Freely diffusible mediators that interact directly with CRM objects; can both accept and donate electrons or protons, forming part of the redox interface. |
| **Support networks** | Thioredoxin system (Trx/TrxR/NADPH); Glutathione system (GSH/GSSG/GR/NADPH) | Regenerate or sustain frontline enzymes and small-molecule pools; modulate the redox potential landscape of the MRM. |
| **Generators** | NADPH oxidase (NOX) isoforms; Nitric oxide synthase (NOS) isoforms; Mitochondrial electron transport chain (multiple sites of leakage) | Primary sources of CRM objects such as $O_2^{\bullet-}$, $H_2O_2$, and $NO\bullet$. Define boundary conditions for energy and electron entry into the redox field. |
| **Macromolecule radicals** | Lipid peroxyl and alkoxyl radicals (ROO•, RO•); Protein thiyl radicals (RS•) | Bridge between CRM small-molecule chemistry and structural biomolecules. Represent the point of redox contact between molecular structure and biological consequence. |

The MRM is bounded by necessity. Only those proteins or cofactors that directly mediate, generate, or regenerate CRM objects are included. Broader metabolic or regulatory components (e.g., iron-binding proteins) are intentionally excluded to preserve structural clarity and interpretability.

Any molecular entity outside of the CRM or MRM is part of the redoxome. The redoxome encompasses all molecules that can theoretically interface with the CRM–MRM via physically admissible transformations. Consequently, the redoxome is necessarily broad: virtually every molecule in the cell qualifies as a potential redoxome object.

For example, any molecule capable of participating in a morphism with hydroxyl radical (•OH)—whether a protein, lipid, or nucleic acid—belongs to the redoxome, since it represents a valid reaction partner [46]. The redoxome also includes molecules that can indirectly influence CRM–MRM activity. For example, an assembly factor for mitochondrial complex I, can theoretically modify the generation of superoxide anion by controlling formation of the catalytic complex [47].

**2.2.4 Example**
The structure of the redox system network emerges as a **tiered** triangle comprising the CRM, the MRM, and the redoxome (**Fig. 2A-B**). Physically admissible transformations can occur both within and across these network modules. For example, hydrogen peroxide, as a CRM object, may undergo reaction with a catalytic cysteine thiolate in peroxiredoxin-2 within the MRM, or with a protein in the redoxome such as PTEN (**Fig. 2C**). In both cases, distinct cysteine proteoforms are generated [48–52], but the transformation context—and thus the functional interpretation—differs.

The MRM acts as a **non-holonomic constraint** on the CRM and, by extension, on the accessible state space of the redoxome. Rather than removing states from the CRM, the MRM constrains how flux can traverse the redox network.

For example, superoxide dismutase (SOD) isoforms impose a strong bias on flux emerging from the superoxide anion, rapidly channelling it into hydrogen peroxide and ground-state oxygen. This corresponds to a heavily weighted set of morphisms that dominate the outgoing transitions from the superoxide object. As a result, the effective availability of superoxide anion for alternative transformations—such as reaction with nitric oxide to form peroxynitrite—is drastically reduced.

By constraining which morphisms are dynamically accessible, SOD governs the functional role of superoxide anion without altering its intrinsic identity as a CRM object. Consequently, the reachable state space of the redoxome is also constrained, since only a limited subset of transformations between superoxide anion and redoxome targets (e.g., aconitase) remain active under these conditions.

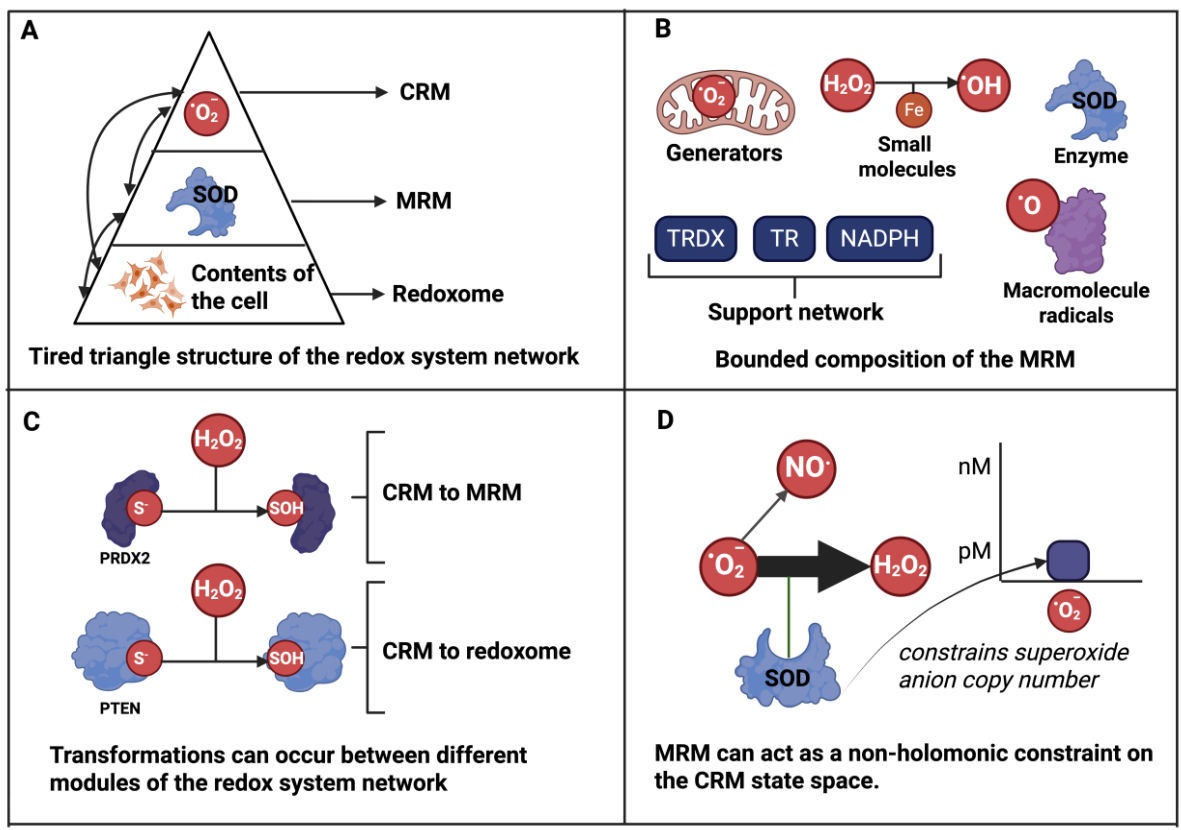

**Figure 2 | Emergence of redox function through modular constraint of flux**

(A) Tiered organisation of the redox system network. The Core Redox Module (CRM) defines the space of admissible small-molecule redox states and transformations. The Modifying Redox Module (MRM) comprises enzymes and cofactors that constrain or mediate CRM transformations. The redoxome encompasses the remaining cellular contents capable of interacting with CRM objects. Function emerges from constrained interactions across these layers.

(B) Bounded composition of the MRM. The MRM consists of a finite set of generators, enzymes, small-molecule mediators, and support networks that directly interface with CRM transformations. This bounded composition defines how flux through the CRM is channelled without altering the underlying state space.

(C) Transformations across network modules. The same CRM object may participate in transformations mediated by the MRM (top) or interact directly with redoxome components (bottom). These alternative transformation contexts yield distinct biological consequences despite originating from the same admissible CRM state.

(D) Constraint of redox trajectories by the MRM. Enzymatic activity (e.g., superoxide dismutase) does not remove states from the CRM but constrains accessible trajectories through the state space by biasing flux into specific transformations. In this sense, the MRM imposes non-holonomic constraints on the realised redox dynamics.

### 2.2.5 Interpretation: Why function cannot be localised to molecules

This construction has immediate biological consequences. Because function arises from relational embedding, no molecule is inherently "functional" or "dysfunctional" in isolation. A molecule such as the superoxide anion acquires meaning only through its participation in specific transformation pathways and through its interaction with enzymes, buffers, and targets.

For example, hydrogen peroxide may be rapidly metabolised by peroxiredoxins or diverted into cysteine-based regulatory pathways. These outcomes do not reflect different identities of hydrogen peroxide itself, but different network embeddings—different distributions of flux across the CRM–MRM–redoxome system.

This perspective resolves longstanding ambiguities in redox biology. Terms such as "oxidant", "antioxidant", or "second messenger" are revealed not as intrinsic descriptors, but as shorthand for dominant flux patterns under certain conditions. A molecule may act in multiple roles over time or across compartments, without contradiction, because function is defined by context rather than identity.

By formalising function as an emergent property of network embedding, this framework separates physical possibility (defined by the CRM) from biological meaning (defined by network configuration). Function therefore follows structure but is not reducible to it. This relational view provides a principled foundation for interpreting redox biology across scales, from local signalling modules to whole-cell redox states [53], and prepares the ground for a dynamic treatment in which function evolves as network fluxes change over time.

### 2.3. Dynamics — Time-Dependent Evolution of Redox State
### 2.3.1 Claim: Redox function is inherently time-dependent

Redox biology is not static. Even when the structural connectivity of a redox system remains unchanged, the rates, magnitudes, and directions of redox transformations vary continuously in time. Molecular abundances fluctuate, enzymatic activities are modulated, and environmental constraints shift. As a consequence, redox function cannot be defined by a single state or snapshot, but only by how redox states evolve.

While the previous section established function as an emergent property of network embedding, that embedding itself is dynamic. The same redox network can realise distinct functional outcomes at different times, not because new transformations become possible, but because existing transformations are weighted differently.

We therefore claim that redox biology must be understood as a time-evolving system, in which structure constrains dynamics, but dynamics determine realised function.

### 2.3.2 Construction: Dynamics as flux evolution on a fixed network

To formalise redox dynamics, we treat the CRM–MRM–redoxome network as structurally fixed over the timescale of interest, while allowing the flux weights associated with its transformations to vary in time. Each transformation is assigned a non-negative, time-dependent weight that represents the effective rate or occupancy of that transformation under prevailing conditions.

At any timepoint, the instantaneous redox state of the system is determined by the current distribution of flux across the network. As these fluxes change, the redox state deforms accordingly. Dynamics are therefore represented as trajectories through the redox state space induced by continuous evolution of flux weights.

Formally, this corresponds to defining a dynamical law on the space of admissible flux configurations, subject to physical constraints such as non-negativity, boundedness, and conservation. The specific form of the dynamical law—whether deterministic, stochastic, or piecewise-defined—is context-dependent and not fixed *a priori*. The same structure-to-function mapping applies at every instant, yielding a time-indexed family of redox states and biological readouts.

The mathematical formulation of these dynamics, including existence, regularity, and invariance properties is provided (**Supplemental Note 3**). The key point is that: redox dynamics arise from time-dependent reweighting of an invariant transformation structure.

### 2.3.3 Example

To illustrate redox dynamics within this framework, we consider a canonical subsystem involving hydrogen peroxide, peroxiredoxins (PRDX), thioredoxin (TDX), and cysteine-containing proteoforms in the redoxome (**Fig. 3A**). Hydrogen peroxide is treated as a CRM object that participates in multiple physically admissible morphisms, including enzymatic reduction via PRDX, regeneration via TDX, and alternative reactions with redoxome components [54].

At any timepoint, the realised redox state of this subsystem is determined by how electron flux derived from hydrogen peroxide is partitioned across these competing transformations. Suppose that at an initial timepoint $t_0$, 95% of the flux is channelled through PRDX-mediated metabolism, with only 5% directed toward cysteine oxidation in the redoxome. Under this partitioning, the dominant functional outcome of the system is to act as an antioxidant by metabolising hydrogen peroxide (**Fig. 3B**).

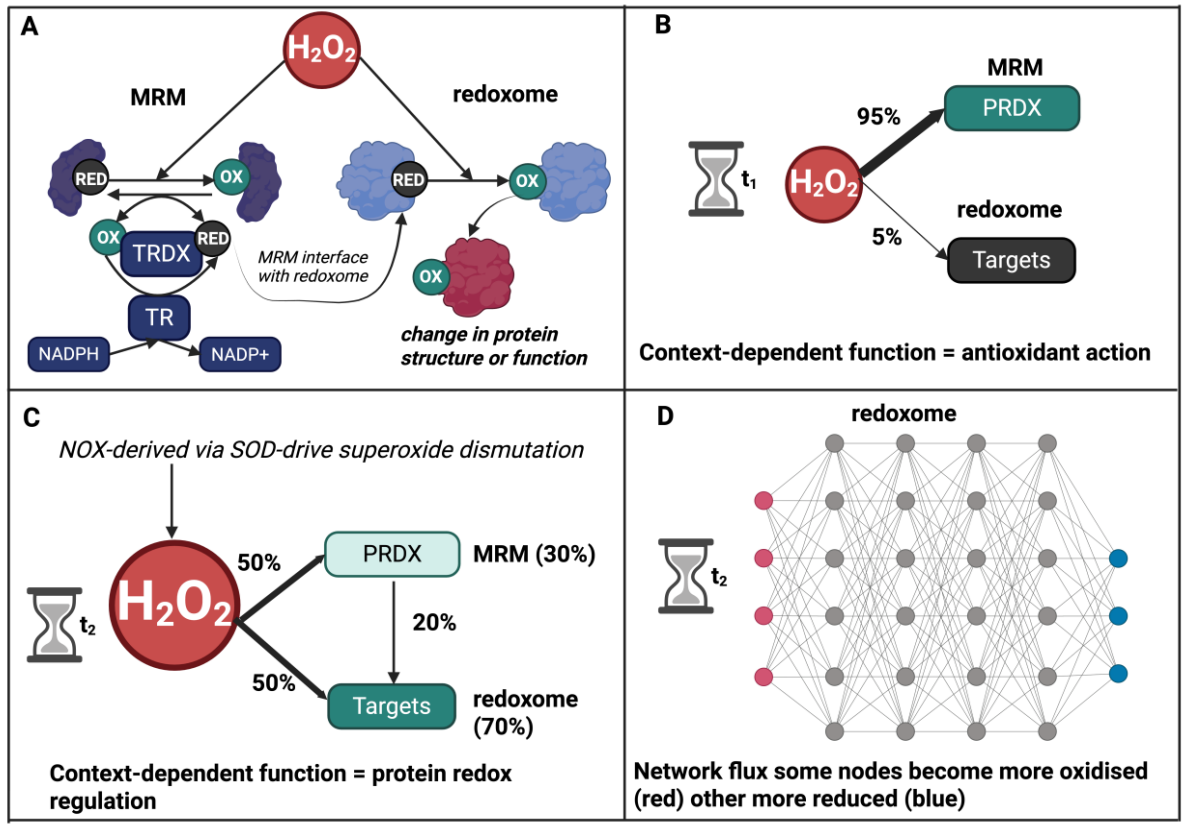

**Figure 3 | Redox dynamics arise from time-dependent redistribution of flux on a fixed network.**
(A) A canonical hydrogen peroxide–centred subsystem spanning the Core Redox Module (CRM), Modifying Redox Module (MRM), and redoxome. Hydrogen peroxide ($H_2O_2$) participates in multiple physically admissible transformations, including enzymatic metabolism via peroxiredoxins (PRDX) and oxidation of redoxome protein targets, with regeneration mediated by the thioredoxin system.
(B) At time $t_1$, flux from $H_2O_2$ is predominantly channelled through the MRM, resulting in rapid detoxification and an emergent antioxidant function.
(C) Following a perturbation that elevates $H_2O_2$ production (e.g., NOX activity with SOD-driven dismutation), the same network structure persists but flux is redistributed at time $t_2$, increasing coupling to redoxome targets and enabling protein redox regulation.
(D) The resulting change in flux distribution alters the redox state of downstream nodes without introducing new molecular entities or reactions. Function emerges from dynamic reweighting of existing morphisms, not from intrinsic molecular identity.

Now consider a perturbation that increases hydrogen peroxide levels. For example, via enhanced NADPH oxidase-driven superoxide anion generation and dismutation after growth factor stimulation [55–57]. At a later timepoint $t_1$, the same network structure persists, but the flux distribution is altered: 70% hydrogen peroxide-derived electrons are now channelled into cysteine oxidation pathways within the redoxome both directly and indirectly via PRDX directed redox relays [58]. Under this reweighting, cysteine redox regulation emerges as a dominant function, despite no change in the underlying set of admissible transformations (**Fig. 3C-D**).

Crucially, these functional differences arise not from changes in molecular identity, but from time-dependent redistribution of flux across a fixed transformation structure. The same CRM object—hydrogen peroxide—supports diverse functional outcomes depending on how its outgoing morphisms are weighted in time. Hence, the structure is invariant but network activation via transformations is mutable.

**2.3.4 Interpretation: Dynamics as the source of context, adaptability, and memory**

Since flux distributions evolve in time, the functional role of a molecular object is mutable. It does not possess a fixed redox function; rather, its role reflects the instantaneous configuration of the network in which it participates [59–61]. Hence, redox dynamics encode context: function reflects not what is present, but how action is distributed at a given moment.

Dynamics introduce history dependence. Past flux configurations influence future ones by shaping molecular abundances, enzyme states, and resource availability [62–66]. As a result, the redox system can exhibit forms of biochemical memory, where prior exposure or activity alters subsequent responses even in the absence of permanent structural change. Such behaviour cannot be captured by static descriptions of redox state but emerges naturally when redox biology is treated as a dynamical system [67].

By extending the structure–function framework into time, dynamics transform redox biology from a catalogue of reactions into a living process. Structure defines what is possible; dynamics determine what is realised and when. This temporal dimension prepares the ground for a spatial extension, where redox dynamics are no longer confined to abstract networks but unfold across physical space.

**2.4. Geometry — Spatial Embedding of Redox Dynamics**

**2.4.1 Claim: Redox dynamics are inherently local and spatially constrained**

Redox transformations occur in space. Each one is a local interaction between molecular objects embedded in a physical environment, constrained by proximity, diffusion, and field-mediated coupling [68–70]. No transformation can influence another without a spatially continuous chain of intermediates or fields connecting them. Locality therefore enforces causality.

Hence, dynamics unfold across spatially distributed molecular environments—organelles, membranes, nanodomains, and tissues—each with its own constraints and interaction neighbourhoods [71].

We therefore claim that redox biology must be described as a spatially embedded dynamical system.

**2.4.2 Construction: The redox system as a spatially distributed field of local algebras**

To formalise spatial organisation, we embed the CRM–MRM–redoxome into a physical domain $\Omega \subset \mathbb{R}^3$, representing a cellular or tissue volume. The domain is partitioned into finite nanodomains, each small enough that redox transformations within it may be treated as locally constrained and environmentally conditioned.

At each location $x \in \Omega$, the system is endowed with a local redox algebra: a weighted transformation graph with the same categorical structure defined previously, but carrying its own time-dependent flux configuration. Each spatial location therefore realises a local instance of the redox system, evolving under local biochemical conditions.

Coupling between neighbouring locations arises through physically admissible transport processes—including molecular diffusion, charge displacement, and field-mediated interactions—which redistribute redox flux across space. These couplings are local and continuous, enforcing spatial causality.

The full mathematical formulation, including reaction–diffusion structure and geometric properties, is provided (**Supplemental Note 4**). The key point is that: there is a spatially distributed redox field, mapping from space and time to redox state, whose evolution is jointly governed by local reaction dynamics and spatial transport.

### 2.4.3 Composition

By construction, the CRM–MRM–redoxome composition of a physical domain is heterogeneous across both space and time. Even when the spatial coordinates defining a domain are held fixed, the molecular composition at that location can vary as redox transformations proceed.

Crucially, this variability involves changes in molecular abundance and identity. For example, when a cysteine residue acts on hydrogen peroxide to form a sulfenic acid, the protein transitions to a distinct cysteine proteoform. This identity change alters the local composition of the domain through explicit object-level transformations encoded in reactant-product couplings.

The physical environment within each spatial domain is itself variable. Local electromagnetic fields, dielectric properties, membrane potentials, and molecular crowding can differ substantially across compartments and over time [72]. For example, a cysteine thiolate positioned proximal to the inner mitochondrial membrane experiences markedly different local electromagnetic field strengths depending on the state of chemiosmotic respiration [73,74]. These field-level variations modulate reaction kinetics and accessibility without altering the underlying set of admissible transformations [68].

As a result, the spatial redox organisation of the cell is best understood as a dynamically evolving mosaic: a heterogeneous patchwork of local redox environments whose composition and functional potential change continuously as molecular identities and fluxes evolve over time.

### 2.4.4 Example

To illustrate spatially embedded redox dynamics, we consider hydrogen peroxide as a CRM object distributed across multiple spatial domains within a cell. Although the admissible transformations of hydrogen peroxide are fixed by the CRM, the realised redox function depends on local embedding and flux distribution.

At a given time point, distinct spatial domains may exhibit different local redox states despite sharing the same underlying transformation structure (**Fig. 4A**). These differences arise from spatially varying flux weights imposed by local enzyme abundance, molecular crowding, and field conditions. As a result, the same molecular species can support divergent redox outcomes in different regions of the cell.

*A Mathematical Theory of Redox Biology*

Local redox composition also evolves in time. At a fixed spatial location, changes in upstream inputs or regulatory activity deform the local flux distribution, yielding time-dependent redox states without altering the underlying network topology (**Fig. 4B**).

Function emerges from this spatial and temporal organisation. In some domains, hydrogen peroxide flux is preferentially channelled through the MRM, producing antioxidant turnover. In others, alternative morphisms dominate, enabling oxidative damage or signalling interactions within the redoxome [75] (**Fig. 4C**). These functional differences do not reflect changes in molecular identity but differences in local flux allocation.

Taken together, these dynamics define a spatially distributed redox field, in which local redox states vary continuously across space and time (**Fig. 4D**). Biological function arises from this field-level organisation rather than from isolated molecules or reactions, reinforcing the view of redox biology as a spatially embedded, dynamical system.

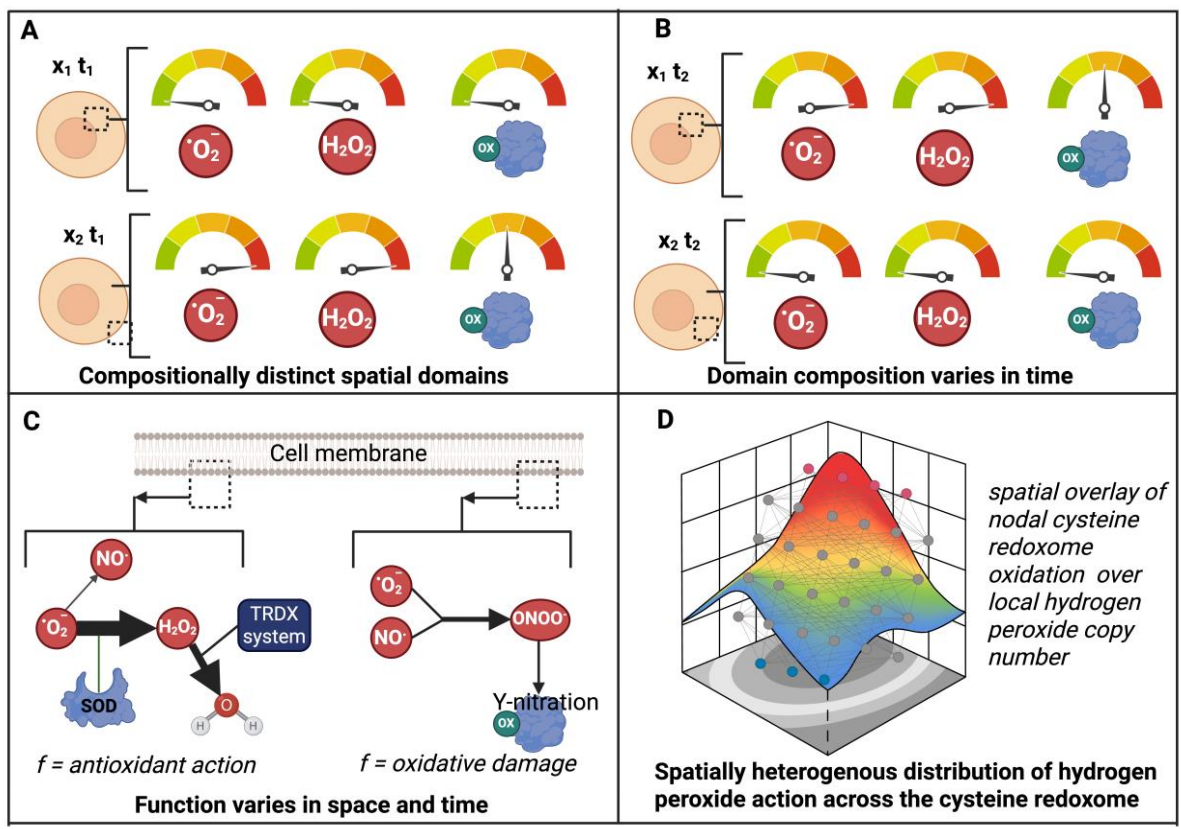

**Figure 4 | Spatial embedding of redox dynamics and emergent function**
(A) Compositional heterogeneity across space. Distinct spatial domains ($x_1$, $x_2$) carry locally instantiated redox systems with identical algebraic structure but different flux distributions across CRM, MRM, and redoxome transformations at the same time point ($t_1$). Local redox state is determined by the relative weighting of admissible transformations rather than by molecular identity alone.
(B) Temporal evolution of local redox composition. At a fixed spatial location, flux distributions evolve over time ($t_1 \rightarrow t_2$), deforming the local redox state without altering the underlying transformation structure. This illustrates that redox state and function are inherently time-dependent.
(C) Spatially localised emergence of function. The same CRM object (e.g. hydrogen peroxide) participates in different dominant transformation pathways depending on spatial embedding and local constraints. In one domain, flux is channelled through the MRM to yield antioxidant metabolism; in another, alternate morphisms dominate, leading to oxidative damage. Some morphisms involving free radical objects like the nitrogen dioxide radical that lead to tyrosine nitration are omitted for visual clarity. Function therefore varies across space even under identical chemical possibilities.

(D) Field-level view of spatial redox organisation. Local redox states form a spatially heterogeneous field, illustrated here as the overlay between nodal cysteine redoxome oxidation and local hydrogen peroxide copy number. Biological outcomes emerge from the coupled evolution of flux distributions across space and time rather than from intrinsic molecular properties.

### 2.4.5 Interpretation: Geometry as the substrate of redox organisation and meaning

Embedding redox dynamics in space transforms the theory from an abstract network description into a field theory of biochemical organisation. Each spatial region behaves as a local redox logic unit, governed by the same compositional rules but operating under distinct constraints and histories.

From this geometric perspective, biological phenomena acquire a natural interpretation. Gradients of redox state correspond to regulatory signals; spatial coherence reflects coordinated activity; compartmentalisation emerges as geometric restriction of flux. Redox "waves" are not metaphorical, but literal propagating deformations of the redox field.

Geometry unifies structure, function, and dynamics. Structure defines the local algebra of possible transformations; dynamics describe how flux evolves in time; geometry determines how these evolving local systems interact across space. Biological meaning emerges from their coupling: function is realised when spatially distributed redox dynamics organise coherently.

In this framework, the living system is governed by the continuous evolution of a spatially extended redox field. Every region of space contains its own redox state and its own history, yet remains causally connected to the whole.

### 3. Conclusion

We have presented a mathematical theory of redox biology that resolves long-standing conceptual ambiguities by treating redox systems as structured, compositional, dynamical, and spatially embedded objects. This work defines a finite redox state space, a principled structure-to-function mapping, time-dependent flux dynamics, and a geometric embedding that enforces locality and causality. Together, these elements constitute a working theory: one that is internally consistent, physically constrained, and operational.

Emergent redox behaviour—such as context dependence, hysteresis, memory, and nonlinearity—does not require additional assumptions. These phenomena arise naturally from bounded state spaces [76], entropy producing reaction networks [77], and flux-driven dynamics on structured networks. Redox function is therefore neither intrinsic nor static, but an evolving property of network embedding across space and time.

This framework provides a predictive interpretative basis. By constraining what redox states and trajectories are admissible, it enables principled reasoning about possible outcomes, transitions, and responses to perturbation. In this sense, the theory provides a rubric for interpreting experimental and modelling data, as well as, a substrate for formulating and testing novel hypotheses. Redox biology is recast as a geometric, dynamical process—one in which biological behaviour emerges from the lawful organisation of biochemical transformation under the living biophysics of the cell.

# Supplemental Notes

# Supplemental Note 1: Formal definition of the Core Redox Module (CRM)

### Defining the CRM

Let the **Core Redox Module (CRM)** be defined as a **small symmetric monoidal category** together with a bounded occupancy structure:

$$\mathcal{C}_{\text{CRM}} = (\text{Obj}(\mathcal{C}), \text{Hom}(\mathcal{C}), \circ, \otimes, I, \alpha, \lambda, \rho, \sigma), \Gamma: \mathbb{N}^{\text{Obj}(\mathcal{C})} \to \mathbb{N}^{\text{Obj}(\mathcal{C})}.$$

Here, $\mathcal{C}_{\text{CRM}}$ encodes **species-level structure**—which molecular states exist and which transformations are physically admissible—while $\Gamma$ encodes **state-space occupancy**, i.e. how many copies of each molecular state may exist within a bounded physical domain.

This separation distinguishes **structural possibility** from **population realisation**.

### Objects: discrete molecular state classes

Each object $A \in \text{Obj}(\mathcal{C})$ represents a **coarse-grained molecular state class**, such as
$O_2$,
$O_2^{\cdot-}$,
$H_2O_2$,
$NO^{\cdot}$,
or
$ONOO^{-}$.

Objects are defined as:

$$A := (\text{comp}(A), s(A)),$$

where:

- $\text{comp}(A)$ denotes elemental composition;
- $s(A)$ is a **state descriptor** capturing physically relevant degrees of freedom.

### State descriptor (coarse-grained)

To avoid literal electron bookkeeping, the state descriptor is defined as:

$$s(A) = (q(A), \pi(A), \varepsilon(A), \chi(A), \dots),$$

where:

- $q(A)$: net charge (integer);
- $\pi(A)$: protonation / acid–base microstate class (finite set);
- $\varepsilon(A)$: excitation, spin, or electronic configuration class (finite set);
- $\chi(A)$: optional **redox-equivalence label** (e.g. oxidation-state class or electron-accepting capacity), used for metrics and comparison, not ontology.

**Interpretation.**
The descriptor $s(A)$ is a **coarse-graining** of the underlying quantum–statistical microstate. It does *not* specify where individual electrons "are"; rather, it encodes the minimal physically meaningful information required to define admissible transformations and compositional reasoning.

Objects may belong to overlapping molecular families via membership maps, e.g.

$$\mathrm{ONOO}^- \in \Omega_O \cap \Omega_N,$$

allowing ROS/RNS/RSS groupings without imposing exclusivity.

## Morphisms: physically admissible transformations

Each morphism $f: A \to B$ denotes a **physically realisable transformation** between coarse molecular states:

$$f := (\tau(f), \Delta s(f), \kappa(f)),$$

where:

- $\tau(f)$ specifies the transformation type (electron transfer, proton transfer, excitation, recombination, abstraction, etc.);
- $\Delta s(f)$ is the induced change in state descriptor;
- $\kappa(f)$ is an optional kinetic or energetic annotation (rate class, catalyst dependence, barrier class).

**Admissibility axiom**

All morphisms are restricted to transformations **admissible under quantum mechanics, statistical mechanics, and thermodynamics**, respecting conservation laws, feasible state transitions, and lawful coupling to external degrees of freedom.

This axiom ensures the CRM encodes only physically meaningful transformations.

**Internal and external morphisms**

- **Internal morphisms** (closed-system):

$$f: A \to B$$

- **External morphisms** (open-system coupling):

$$f: (A \otimes X) \to B$$

where $X$ represents an external object such as a photon ($h\nu$), metal ion, membrane potential surrogate, or field interaction.

## Monoidal structure: composition and concurrency

The tensor product

$$\otimes: \mathcal{C} \times \mathcal{C} \to \mathcal{C}$$

encodes **parallel composition**.

- On objects: $A \otimes B$ denotes joint presence of $A$ and $B$ in a composite local system.
- On morphisms:

$$f_1 \otimes f_2: (A_1 \otimes A_2) \to (B_1 \otimes B_2)$$

represents concurrent, independent transformations.

The unit object $I$ represents the inert background.
Natural isomorphisms $\alpha, \lambda, \rho, \sigma$ satisfy Mac Lane coherence conditions.

Sequential composition $g \circ f$ represents reaction chains; monoidal composition represents concurrency. Together, they encode the **compositional algebra of redox chemistry**.

## Bounded occupancy: population state space

To represent statements such as "superoxide is present or absent, and if present exists in $1, \ldots, N$ copies", we introduce a bounded population structure.

Let a physical domain $V$ (e.g. voxel, compartment, or cell) be fixed.
For each object $A$, define a maximum physically allowable count $N_A(V) \in \mathbb{N}$, determined by packing, solubility, or physical constraints.

The **CRM occupancy space** is:

$$\mathcal{M}_V = \prod_{A \in \mathrm{Obj}(\mathcal{C})} \{0, 1, \ldots, N_A(V)\}.$$

A system state is a marking $m \in \mathcal{M}_V$, where $m(A)$ denotes the copy number of object $A$ in domain $V$.

A Mathematical Theory of Redox Biology

All bounds are domain-relative and may differ across compartments or spatial regions.

## Morphism action on occupancies

Each morphism induces a stoichiometric update on markings:

$$m \mapsto m' = \text{clip}_{\mathcal{M}_V}(m + \nu_f),$$

where:

- $\nu_f \in \mathbb{Z}^{\text{Obj}(\mathcal{C})}$ is the stoichiometric incidence vector (negative for consumption, positive for production);
- $\text{clip}_{\mathcal{M}_V}$ enforces bounds $0 \leq m'(A) \leq N_A(V)$.

This induces a **Petri-net–like or reaction-network operational semantics**, without reducing the CRM ontology to a Petri net.

The map Γ acts on markings by applying these admissible morphism-induced updates, defining population evolution within the bounded state space.

## Metrics on the state space

Define per-channel distances over admissible transformation paths:

$$\delta_u(A, B) = \min_{\text{paths } A \to B} \left| \sum_i \Delta s_u(f_i) \right|, u \in \{q, \pi, \varepsilon, \chi\}.$$

A combined metric may be defined as:

$$\delta(A, B) = \sqrt{\omega_q \delta_q^2 + \omega_\pi \delta_\pi^2 + \omega_\varepsilon \delta_\varepsilon^2 + \omega_\chi \delta_\chi^2}.$$

These metrics quantify **transformation magnitude within the admissible state space**. They are **not assumed to correspond directly** to thermodynamic free energy or kinetic rates.

## Interpretation

The CRM defines a **finite, compositional redox state space** whose elements are physically meaningful molecular states and whose morphisms encode lawful transformations. Objects are neutral with respect to biological role; morphisms describe what can occur, not what *should* occur.

Population, dynamics, function, and geometry arise only when this neutral algebra is embedded into broader biological context. The CRM therefore provides the structural foundation upon which function, dynamics, and spatial organisation are constructed in the main theory.

# Supplemental Note 2: Structure–Function Mapping in Redox Systems

**Defining the composite redox interaction graph**

Let

$$G = (V, E, s, t, \tau)$$

be a **finite typed directed graph** representing the combined **CRM–MRM–redoxome** system.

- $V$ is the set of molecular objects (nodes), including small molecules (CRM), modifying enzymes and cofactors (MRM), and all other redox-interacting species (redoxome).
- $E$ is the set of biochemical transformations (edges).
- $s, t: E \to V$ are source and target maps.
- $\tau(e)$ assigns each edge a **channel type** (e.g. electron transfer, proton transfer, excitation, covalent modification).

This graph encodes **which interactions are possible**, but not yet **how strongly** they are expressed.

**Flux algebra: state on structure**

To represent how the redox network is *used* under specific conditions, we assign **effective flux weights**:

$$\omega: E \to \mathbb{R}_{\geq 0}.$$

Each $\omega(e)$ represents the **effective occupancy or throughput** of transformation $e$ over a defined observation window (e.g. steady state, time interval $t_0 \to t_1$). These weights are coarse-grained summaries of biochemical activity and do not imply microscopic current flow.

Define the **incidence matrix**:

$$D \in \mathbb{Z}^{V \times E}, \quad D_{v,e} = \begin{cases} +1 & \text{if } t(e) = v \\ -1 & \text{if } s(e) = v \\ 0 & \text{otherwise.} \end{cases}$$

The **net node flux** is:

$$\mu = D\omega \in \mathbb{R}^V,$$

which represents the imbalance of incoming and outgoing transformations at each molecular node.

## Molecular attribute map

Each node $v \in V$ carries a vector of **physical attributes**:

$$q(v) = (q_1, q_2, \ldots, q_p) \in \mathbb{R}^p,$$

encoding coarse-grained properties such as redox-equivalent capacity, proton exchange potential, or energy coupling class.

Collect these into a matrix:

$$Q \in \mathbb{R}^{p \times V}.$$

**Interpretation.**
The map $Q$ does **not** represent a literal accounting of electrons or protons. It defines a **projection from network structure to chemically interpretable coordinates**, suitable for aggregation and comparison.

## Global redox state

The **global redox state** of the system is defined as the linear functional:

$$R_{\text{global}} = QD\omega \in \mathbb{R}^p.$$

This vector summarises the aggregate balance of redox-relevant channels across the entire CRM–MRM–redoxome network for the chosen observation window.

## Local redox submaps and naturality

For any induced subgraph $H \subseteq G$ (e.g. an $H_2O_2$-centred subsystem), let $P_H$ be the node-selection operator.

Define restricted operators:

$$Q_H = QP_H^\top, \quad D_H = P_H D.$$

The **local redox state** is:

$$R_H = Q_H D_H \omega.$$

Naturality condition:
If $H \subseteq K \subseteq G$, then

$$R_H = P_H R_K,$$

ensuring that local states are consistent projections of larger contexts. This guarantees **modularity** and **scale coherence**.

## Functorial structure

Define the **redox state functor**:

$$\text{Redox}: \textbf{WGraph} \rightarrow \textbf{State}, \text{Redox}(G, \omega, Q) = (\mathbb{R}^p, \omega \mapsto QD\omega),$$

where:

- **WGraph** is the category of weighted directed graphs,
- **State** is the category of finite-dimensional real vector spaces and linear maps.

The functor is **monoidal**:

$$\text{Redox}(G_1 \sqcup G_2, \omega_1 \sqcup \omega_2) = \text{Redox}(G_1, \omega_1) \oplus \text{Redox}(G_2, \omega_2),$$

reflecting independent composition of disjoint subsystems.

## Composite structure of redox biology

The full system decomposes as:

$$G = G_{\text{CRM}} \sqcup G_{\text{MRM}} \sqcup G_{\text{Redoxome}},$$

with cross-edges encoding catalytic, regulatory, or destructive interactions.

The same algebra applies uniformly to:

- the full system,
- any subgraph,
- or any intermediate modular assembly.

## Structure-to-function mapping

Biological meaning arises only when the redox state is composed with a **readout map**:

$$\text{Func} = \text{Read} \circ \text{Redox}: (G, \omega, Q) \mapsto \text{biological output}.$$

Here:

- Read: $\mathbb{R}^p \to Y$ maps chemical state vectors to biological variables (e.g. enzyme activity, signalling probability, metabolic flux),
- $Y$ is an application-dependent biological space.

**Interpretation.**
Function is not an intrinsic property of any molecule or reaction. It emerges from:

1. network structure (what can interact),
2. flux distribution (what is active),
3. and biological readout (what is sensed or acted upon).

This formalism makes explicit how **structure gives rise to function** without assigning fixed roles to molecular identities. The same chemistry can support signalling, buffering, or damage depending solely on how flux is routed through the network.

## Conceptual summary

The structure–function mapping treats redox biology as a **weighted relational system**, not as a collection of labelled molecules. The redox state is a projection of network activity, and biological function is a further projection of that state into organismal behaviour.

This framework preserves neutrality at the chemical level while allowing meaning to emerge at higher levels—preparing the ground for temporal dynamics and spatial embedding in subsequent sections.

# Supplemental Note 3: Dynamics of Redox Systems

## Motivation

Structure–function mappings describe redox systems at a fixed observation window. Biological systems, however, are not static: molecular fluxes change over time in response to internal regulation, external stimuli, and resource constraints. As a result, the redox state of a system is inherently time-dependent.

A mathematical theory of redox biology must therefore extend the structure–function formalism to describe **how flux distributions evolve**, while preserving the compositional and modular properties of the underlying network.

## Dynamic variables

Let

$$G = (V, E, s, t, \tau)$$

be the fixed CRM–MRM–redoxome interaction graph defined previously.

Dynamics are introduced by allowing the edge weights (fluxes) to vary with time:

$$\omega_t: E \to \mathbb{R}_{\geq 0}, t \in [0, T].$$

The graph topology and molecular attribute map $Q$ remain fixed on the time scale considered; only the **distribution of flux across admissible transformations** evolves.

## Time-dependent redox state

At each time $t$, the instantaneous redox state is:

$$R_t = QD\omega_t \in \mathbb{R}^p,$$

with local states for any induced subgraph $H \subseteq G$:

$$R_{H,t} = Q_H D_H \omega_t.$$

Thus, redox dynamics correspond to a **trajectory through redox state space**, induced by changes in flux allocation rather than by changes in molecular identity.

## Dynamical law on flux weights

We define a general evolution law on flux weights:

$$\dot{\omega}_t = F(\omega_t, \mu_t, \Theta),$$

where:

- $F: \mathbb{R}_{\geq 0}^E \times U \times \Theta \to \mathbb{R}^E$ is a vector field;
- $\mu_t \in U$ represents time-dependent control inputs (e.g. nutrient availability, stress, signalling cues);
- $\Theta$ denotes fixed system parameters (e.g. enzyme abundance classes, compartmental constraints).

### Regularity assumptions

To ensure well-posedness, we assume:

1. **Carathéodory conditions**: measurable in $t$, continuous in $\omega$;
2. **Local Lipschitz continuity** in $\omega$;
3. **Forward invariance** of the non-negative cone $\mathbb{R}_{\geq 0}^E$.

These assumptions guarantee existence and uniqueness of solutions on finite intervals.

## Induced dynamics on redox state

The time derivative of the redox state follows directly:

$$\dot{R}_t = QD\dot{\omega}_t = QDF(\omega_t, \mu_t, \Theta).$$

Thus, the same linear structure that defined redox state at fixed time also governs its temporal evolution.

This yields a **structure-preserving dynamic system**: the algebra does not change over time; only its instantiation does.

**Functorial time evolution**

Define the time-indexed redox functor:

$$\text{Redox}_t : (G, \omega_t, Q) \mapsto R_t.$$

Then $\{\text{Redox}_t\}_{t \in [0,T]}$ defines a **smooth functor flow**, mapping a continuous trajectory in flux space to a continuous trajectory in state space.

Naturality under restriction is preserved:

If $H \subseteq G$, then

$$R_{H,t} = P_H R_t \forall t,$$

so local subsystems inherit consistent dynamics from the global system.

**Biological readout under dynamics**

Let

$$\text{Read} : \mathbb{R}^p \to Y$$

be a (typically smooth) readout map into a biological variable space $Y$.

Then the time-dependent biological output is:

$$y_t = \text{Read}(R_t), \dot{y}_t = D\text{Read}(R_t)\dot{R}_t.$$

This expresses biological behaviour as a **derived observable** of redox dynamics, not as a primary variable.

**Interpretation**

Redox dynamics are not driven by changes in what reactions are possible, but by changes in **which reactions are utilised and to what extent**. The same structural network can support qualitatively different behaviours depending on how flux is distributed over time.

In this framework:

- **Structure** defines the admissible transformation space;
- **Flux dynamics** determine the active pathways;
- **Redox state deformation** encodes system-level change;
- **Biological function** emerges as a time-dependent readout.

Because past flux configurations influence future states through $\omega_t$, the system naturally supports history dependence, adaptation, and memory-like effects without introducing additional state variables.

## Conceptual summary

Dynamics extend the redox algebra into time by allowing flux weights to evolve under lawful constraints. The resulting theory treats redox biology as a **living system of constrained flows**: a structured network whose meaning changes as electrons, protons, and energy are redistributed.

This dynamic formalism prepares the ground for spatial embedding, where flux evolution is further constrained by locality and transport, yielding a geometric field description of redox biology.

# Supplemental Note 3 (extended): Control Structure and Non-holonomic Constraints

### Control-theoretic interpretation

The dynamical law

$$\dot{\omega}_t = F(\omega_t, \mu_t, \Theta)$$

admits a natural interpretation within control theory.

Here, the flux weights $\omega_t$ define the **state variables** of the redox system, while the control inputs $\mu_t$ represent externally or internally modulated signals that bias flux redistribution. These controls may correspond to changes in enzyme activity, substrate availability, redox cofactor supply (e.g. NADPH), compartmental conditions, or signalling inputs.

Under this interpretation, redox biology is modelled as a **controlled dynamical system** evolving on a constrained state space:

$$\omega_t \in \Omega \subset \mathbb{R}^E_{\geq 0},$$

with admissible trajectories determined by both structural constraints (graph topology) and control constraints (available modulation channels).

Importantly, the control inputs $\mu_t$ do not introduce new transformations. They act by **reweighting existing morphisms**, selectively amplifying, suppressing, or rerouting flux through the pre-defined redox network.

## The Modifying Redox Module as a non-holonomic constraint

Within this framework, the Modifying Redox Module (MRM) plays a precise mathematical role.

The CRM defines the **configuration space** of admissible redox transformations: what molecular transitions are physically possible. The MRM does not alter this space. Instead, it constrains **how the system may move through it over time**.

Formally, the MRM imposes **non-holonomic constraints** on the flux dynamics. That is, it restricts the allowable directions of motion in flux space without defining a potential function whose gradient determines the dynamics.

Concretely:

- The presence, abundance, and kinetic properties of enzymes constrain which combinations of flux changes are achievable;
- These constraints are path-dependent and history-sensitive;
- They cannot, in general, be integrated into a scalar state function.

Thus, while the CRM defines the admissible morphisms, the MRM defines the **admissible trajectories** through the space of flux configurations.

This mirrors classical non-holonomic systems in mechanics, where constraints limit velocity vectors but not configuration variables. Here, the redox system may occupy the same instantaneous state $R_t$ via distinct flux histories, with different biological consequences.

## Controlled accessibility and biological meaning

From a control perspective, biological regulation corresponds to shaping the **reachable set** of redox states over time.

Two systems with identical CRM structure and identical instantaneous redox state may differ in:

- which future states are accessible,
- how rapidly transitions can occur,
- whether certain trajectories are forbidden.

These differences arise entirely from MRM-imposed constraints on flux evolution.

This framing explains why redox function is:

- **context-dependent** (control inputs differ),
- **history-dependent** (non-holonomic paths matter),
- **irreducible to static potentials** (no global energy landscape suffices).

Biological meaning, in this view, emerges not from the redox state alone, but from the **controlled motion of the system through redox state space**.

### Interpretation

By making the control structure explicit, the framework clarifies the role of enzymes, cofactors, and regulatory systems as **trajectory-shaping operators** rather than state-defining entities.

The CRM specifies what chemistry is possible.
The MRM specifies how that chemistry may unfold.
Dynamics arise from their interaction.

This establishes redox biology as a controlled, non-holonomic dynamical system on a structured chemical manifold—preparing the ground for spatial embedding and geometric interpretation.

# Supplemental Note 4. Redox Field Formalism

## Domain and local systems

Let

$$\Omega \subset \mathbb{R}^3$$

denote a bounded biological domain (e.g. a cell, organelle, or tissue region), discretised into voxels $V_i$ of characteristic physical scale

$$\ell \sim 0.5 \text{ nm}.$$

Each spatial point $x \in \Omega$ carries a **local redox system**, represented by a weighted directed graph

$$(G_x, \omega_x(t), Q_x),$$

encoding the interactions of the CRM–MRM–redoxome at position $x$ and time $t$.

The graph topology may vary across space to reflect compartmentalisation, molecular localisation, or structural heterogeneity.

## Local redox state

At each point $x$ and time $t$, define the instantaneous local redox state vector

$$R(x,t) = Q_x D_x \omega_x(t) \in \mathbb{R}^p,$$

where:

- $D_x$ is the incidence matrix of $G_x$,
- $Q_x$ maps molecular nodes to physical redox attributes,
- $\omega_x(t)$ are the local flux weights.

This defines the **local algebraic state** of redox activity at each spatial location.

## Redox field

Collectively, the family $\{R(x,t)\}_{x \in \Omega}$ defines a **redox field**

$$R: \Omega \times \mathbb{R}_{\geq 0} \longrightarrow \mathbb{R}^p,$$

assigning to each spatial location and time a structured redox state.

## Spatial flux coupling

Spatial interaction between neighbouring voxels is represented by a flux tensor

$$J(x,t) = -\kappa(x)\, \nabla R(x,t),$$

where $\kappa(x)$ is a (possibly anisotropic) transport tensor encoding effective diffusivity or field-mediated conductance (electronic, protonic, ionic, or mixed).

This term captures transport, propagation, and spatial coupling of redox information.

## Field equation (reaction–diffusion–control form)

The conservation of redox state across space–time is governed by

$$\frac{\partial R(x,t)}{\partial t} = Q_x D_x F\left(\omega_x(t), \mu_x(t), \Theta_x\right) - \nabla \cdot J(x,t),$$

which couples:

- **internal controlled reaction dynamics** (first term),
- **spatial transport and field propagation** (second term).

Here:

- $F$ is the same control-driven vector field defined in Supplemental Note 3,
- $\mu_x(t)$ represents spatially local control inputs,
- $\Theta_x$ encodes local structural or biochemical parameters.

## Geometric structure

Define the redox field map

$$\text{Field}: (x, t) \mapsto (R(x,t), J(x,t)).$$

This map defines a smooth section of a vector bundle over $\Omega$, whose fibers encode the local redox algebra and whose connection (induced by $\kappa$) governs inter-voxel communication.

Each fiber carries its own instance of the CRM–MRM–redoxome structure; spatial coupling deforms how these local algebras interact.

## Spatial functoriality and locality

The system is organised as a **spatial functor**

$$\text{RedoxField}: \text{WGraph}_\Omega \rightarrow \text{Field}(\Omega),$$

assigning to each voxel $x$ a local redox system $\text{Redox}_x$.

Neighbouring systems commute under spatial transport:

$$\text{Redox}_{x+\delta x} - \text{Redox}_x = \nabla_x \text{Redox}_x \cdot \delta x.$$

Thus, locality is respected and causality propagates through continuous spatial morphisms rather than nonlocal shortcuts.

## Interpretation

The spatial embedding transforms the redox system from an abstract dynamical network into a **field theory of biochemical organisation**.

Each voxel is a local algebra of redox transformations, evolving under controlled, non-holonomic flux dynamics. The global redox state emerges from the geometric coupling of these local systems through diffusion and field-mediated transport.

This construction unifies:

- **structure** (local admissible chemistry),
- **function** (flux-weighted connectivity),

- **dynamics** (controlled evolution),
- **geometry** (spatial embedding and propagation).

The resulting redox field $R(x,t)$ provides a continuous, causal representation of chemical information flow through biological space—where gradients encode regulation, spatial coherence encodes communication, and field deformation corresponds to adaptation.